\documentclass[fleqn,twoside]{article}
\usepackage[headings]{espcrc2a}
\usepackage{axodraw}
\usepackage{colordvi}
\newcommand{\nl}{\nonumber\\}

\newcommand{\lpar}{\left(}                            
\newcommand{\rpar}{\right)}

\newcommand{\bq}{\begin{equation}}                    
\newcommand{\eq}{\end{equation}}
\newcommand{\bqa}{\arraycolsep 0.14em\begin{eqnarray}}
\newcommand{\eqa}{\end{eqnarray}}
\newcommand{\ba}[1]{\begin{array}{#1}}
\newcommand{\ea}{\end{array}}
\newcommand{\ben}{\begin{enumerate}}
\newcommand{\een}{\end{enumerate}}
\newcommand{\bei}{\begin{itemize}}
\newcommand{\eei}{\end{itemize}}
\newcommand{\eqn}[1]{Eq.(\ref{#1})}

\newcommand{\tabn}[1]{Tab.~\ref{#1}}

\newcommand{\tabns}[2]{Tabs.~\ref{#1}--\ref{#2}}

\newcommand{\GeV}{\mathrm{GeV}}

\def\Re{\mathop{\operator@font Re}\nolimits}
\def\Im{\mathop{\operator@font Im}\nolimits}

\newcommand{\wb}{W}

\newcommand{\mz}{M_{_Z}}

\newcommand{\mh}{M_{_H}}

\newcommand{\mzs}{M^2_{_Z}}

\newcommand{\gf}{G_{\ssF}}

\newcommand{\li}[2]{\mathrm{Li}_{#1}\lpar\displaystyle{#2}\rpar} 

\newcommand{\MSB}{\overline{MS}}

\newcommand{\srt}{\sqrt{2}}

\newcommand{\Reb}{{\rm{Re}}}
\newcommand{\Imb}{{\rm{Im}}}
\newcommand{\upar}[1]{u}

\newcommand{\ssF}{{\scriptscriptstyle{F}}}
\newcommand{\ssG}{{\scriptscriptstyle{G}}}
\newcommand{\ssH}{{\scriptscriptstyle{H}}}

\newcommand{\ssK}{{\scriptscriptstyle{K}}}

\newcommand{\ssM}{{\scriptscriptstyle{M}}}

\newcommand{\ssO}{{\scriptscriptstyle{O}}}
\newcommand{\ssP}{{\scriptscriptstyle{P}}}
\newcommand{\ssQ}{{\scriptscriptstyle{Q}}}
\newcommand{\ssR}{{\scriptscriptstyle{R}}}
\newcommand{\ssS}{{\scriptscriptstyle{S}}}

\newcommand{\ssV}{{\scriptscriptstyle{V}}}
\newcommand{\ssW}{{\scriptscriptstyle{W}}}

\newcommand{\bqas}{\begin{eqnarray*}}
\newcommand{\eqas}{\end{eqnarray*}}

\def\app#1#2 {{\it Acta. Phys. Pol.} {\bf#1},#2}
\def\cpc#1#2 {{\it Computer Phys. Comm.} {\bf#1},#2}
\def\np#1#2 {{\it Nucl. Phys.} {\bf#1},#2}
\def\pl#1#2 {{\it Phys. Lett.} {\bf#1},#2}
\def\prep#1#2 {{\it Phys. Rep.} {\bf#1},#2}
\def\prev#1#2 {{\it Phys. Rev.} {\bf#1},#2}
\def\prl#1#2 {{\it Phys. Rev. Lett.} {\bf#1},#2}
\def\zp#1#2 {{\it Zeit. Phys.} {\bf#1},#2}
\def\sptp#1#2 {{\it Suppl. Prog. Theor. Phys.} {\bf#1},#2}
\def\mpl#1#2 {{\it Modern Phys. Lett.} {\bf#1},#2}
\def\jetp#1#2 {{\it Sov. Phys. JETP} {\bf#1},#2}
\def\fpj#1#2 {{\it Fortschr. Phys.} {\bf#1},#2}
\def\afp#1#2 {{\it Acta.Phys. Polon.} {\bf#1},#2}
\def\err#1#2 {{\it Erratum} {\bf#1},#2}
\def\ijmp#1#2 {{\it Int. J. Mod. Phys} {\bf#1},#2}
\def\nc#1#2 {{\it Nuovo Cimento} {\bf#1},#2}
\def\ap#1#2 {{\it Ann. Phys.} {\bf#1},#2}
\def\cmp#1#2 {{\it Comm. Math. Phys.} {\bf#1},#2}
\def\el#1#2 {{\it Europhys. Lett.} {\bf#1},#2}
\def\hpa#1#2 {{\it Helv. Phys. Acta} {\bf#1},#2}
\def\yf#1#2 {{\it Yad. Fiz.} {\bf#1},#2}
\def\nim#1#2 {{\it Nucl. Instrum. Meth.} {\bf#1},#2}
\def\spz#1#2 {{\it Sov. Pisma Zhetf} {\bf#1},#2}
\def\jetpl#1#2 {{\it JETP Lett.} {\bf#1},#2}
\def\sjnp#1#2 {{\it Sov. J. Nucl. Phys.} {\bf#1},#2}
\def\ptp#1#2 {{\it Progr. Theor. Phys. (Kyoto)} {\bf#1},#2}
\def\rmp#1#2  {{\it Rev. Mod. Phys.} {\bf#1},#2}
\def\zhetf#1#2 {{\it ZhETF} {\bf#1},#2}
\def\prs#1#2 {{\it Proc. Roy. Soc.} {\bf#1},#2}
\def\phys#1#2 {{\it Physica} {\bf#1},#2}

\def\bfi{\begin{figure}}
\def\efi{\end{figure}}

\newcommand{\dcub}[1]{\int\,dC_{#1}}

\newcommand{\bca}{\ssK}
\newcommand{\bcan}[1]{#1;\ssK}

\newcommand{\LB}{{\cal L}oop{\cal B}ack}
\newcommand{\GS}{{\cal G}raph{\cal S}hot}

\newcommand{\bmid}{\Bigr|}

\newcommand{\cpw}{s_{\ssW}}

\newcommand{\rpw}{\mu^2_{\ssW}}

\usepackage{graphicx}
\usepackage[figuresright]{rotating}

\newcommand{\AmS}{{\protect\the\textfont2
  A\kern-.1667em\lower.5ex\hbox{M}\kern-.125emS}}

\title{
\vspace{-0.8cm}
{\scriptsize{\texttt{DESY 06-121}}}\\
Two Loop QFT in the Making}

\author{Stefano~Actis\address{Deutsches Elektronen-Synchrotron, 
Zeuthen, Germany}, 
Giampiero~Passarino\address[TO]{Dipartimento di Fisica Teorica, Universit\`a 
                                di Torino, Italy\\
                                INFN, Sezione di Torino, Italy}, 
Sandro~Uccirati\addressmark[TO]}
\runauthor{S. Actis, G. Passarino, S. Uccirati}

\begin{document}

\begin{abstract}
Recent developments in the evaluation of two-loop pseudo-observables and 
observables are briefly reviewed. 
\vspace{1pc}
\end{abstract}

\maketitle

\section{Introduction}
The complete strategy to derive theoretical predictions for pseudo-observables
(PO) and observables (O) up to two-loop accuracy is based on the following 
steps: generation and manipulation of diagrams, renormalization, 
semi-numerical evaluation of diagrams. We perform the first step with the help
of the code $\GS$ (FORM 3.1)~\cite{GraphShot} while the last is performed 
using the code $\LB$ (FORTRAN 95)~\cite{LB} which makes extensive use 
of array handling, assignment overloading, vector/recursive functions.
\section{Renormalization and Unitarity}
A keyword in renormalization is {\em counterterm}; they are not
strictly needed but are, nevertheless, very useful when dealing with 
overlapping divergencies. $\GS$ generates all counterterms needed in the
standard model and produces ultraviolet (UV) finite Green functions. Another 
important keyword is {\em skeleton expansion}, meaning that the relevant 
objects in perturbation theory are dressed propagators. With their help we 
want to construct a (finite) renormalization procedure where a) the 
renormalized parameters are real, b) finite renormalization is the result of a 
consistently truncated solution of renormalization equations (RE), c) complex 
poles arise naturally after dressing the propagators, but cutting equations 
remain valid to all orders. For an unstable particle $V$ define
\bq
{\bar \Delta}_{\ssV} = 
\frac{\Delta_{\ssV}}{1 - \Delta_{\ssV}\,\Sigma_{\ssV\ssV}},
\label{dp}
\eq
where $\Sigma$ is the $V$ (skeleton) self-energy. Cutting-equations and 
unitarity of the $S\,$-matrix can be proven: one uses two-loop
${\bar \Delta}$ in tree diagrams, one-loop ${\bar \Delta}$ in one-loop
diagrams and tree propagators in two-loop diagrams.
The proof is due to Veltman~\cite{Veltman:1963th}:
the crucial observation is that ${\bar \Delta}$ satisfies the 
K\"allen - Lehmann representation,
\bq
{\bar \Delta}^{+}_{\ssV}(p^2) = \theta(p_0)\,
\Bigl[ {\bar \Delta}_{\ssV}(p^2)\Bigr]^2\,
2\,\Imb\,\Sigma_{\ssV\ssV}(p^2),
\eq
while, for a stable particle $s$, the pole term shows up as
\bqa
{\bar \Delta}^{+}_{s}(p^2) &=& \theta(p_0)\,
\Bigl[ {\bar \Delta}_{s}(p^2)\Bigr]^2\,
2\,\Imb\,\Sigma_{ss}(p^2) 
\nl
{}&+& 2\,i\,\pi\,\delta(p^2 + m^2_{s}).
\eqa
One then expresses $\Imb\,\Sigma_{\ssV\ssV}$ in terms of cut self-energies, 
repeats ad libidum and derives that contributions from cut lines come
from stable particles only. Consider a toy model with 
\bq
L_{\rm int} = \frac{g}{2}\,\Phi(x)\,\phi^2(x),
\eq
and where $\Phi$ is unstable. We define ${\overline \Delta}_{\Phi}$ and
${\overline \Delta}_{\phi}$ according to \eqn{dp}.
%
An example of the skeleton expansion for self-energies is given in Fig.~1:
$\Imb\,\Sigma_{\phi\phi} \not= 0$ only due to the $3-$particle cut of
diagram b) of Fig.~1 and only diagrams a) and c) are retained in the
expansion; in a) we use ${\overline \Delta}_{\Phi}$, at one-loop accuracy.

In a gauge theory, however, there is a clash between resummation and gauge 
invariance; usually only the complex pole is resummed~\cite{Denner:2005fg}. 
Therefore a one-loop self-energy with ${\overline \Delta}_{\Phi}$ at one-loop 
accuracy is equivalent to the $3$ diagrams of Fig.~2 computed with 
$\Delta_{\Phi}(s_{\ssM})$, where 
\bq
Z_p = \frac{g^2}{16\,\pi^2}
B_0 \lpar - s_{\ssM}\,;\,m\,,\,m\rpar.
\eq
The interplay of gauge parameter independence, Ward - Slavnov - Taylor
identities and unitarity (which is naturally satisfied in the framework of
dressed propagators) requires a more detailed analysis, beyond the scope of
this paper.

To solve REs we need an input parameter set (IPS) including some notion of
$M_{\rm exp}$. In the past on-sell PO have been derived by fitting lineshapes 
from experiments~\cite{Grunewald:2000ju} but we can use an on-shell 
$M_{\ssO\ssS}$ only at one-loop. Beyond this order the correct treatment 
requires introducing complex poles~\cite{Beenakker:1996kn},
$s_{\ssV} = \mu^2_{\ssV} - i\,\gamma_{\ssV}\,\mu_{\ssV}$.

If we want to use available data a transformation is therefore needed:
define pole PO through $\psi = \arctan \Gamma_{\ssO\ssS}/M_{\ssO\ssS}$:
$M_{\ssP} = M_{\ssO\ssS}\,\cos\psi$ and 
$\Gamma_{\ssP} = \Gamma_{\ssO\ssS}\,\sin\psi$.
Once again, a change of strategy is needed since RE change their structure
at two loops. 
It is a new perspective: at one loop one considers $M_{\ssO\ssS}$ as an input
parameter independent of $s_{\ssP}$ and derive $s_{\ssP}$.
At two loop REs are written for real $p_{\ssR}$ and solved in terms of 
(among other things) experimental $s_{\ssP}$.

In the framework of an order-by-order renormalization, $M_{\ssR}$ is 
a real solutions of truncated REs consistently with cutting-equations and 
unitarity.
\subsection{Complex poles}
Here we give an example of the old fashioned one-loop technique:
on-shell masses are input and, having the Higgs boson in mind, we derive 
$s_{\ssH} = \mu^2_{\ssH} - i\,\mu_{\ssH}\,\gamma_{\ssH}$; numerical results 
for $s_{\ssH}$ are shown in \tabn{of}.

The current fashion instead is to extract $s^{\rm exp}_{\ssH} = \mu^2_{\ssH} -
i\,\mu_{\ssH}\,\gamma_{\ssH}$ from data and to derive
$s^{\rm th}_{\ssH} = M^2_{\ssH} - i\,M_{\ssH}\,\Gamma_{\ssH}$.
To see how it works we point out the difference with previous two-loop 
calculations~\cite{Jegerlehner:2003wu} where one starts from
\bq
s_{\ssV} = m^2 - \Sigma_{\ssV\ssV}\lpar s_{\ssV}\,,\,m^2\,,\,\dots\,\rpar,
\eq
and derives the complex pole in terms of the (bare) renormalized mass 
\bq
s_{\ssV} = m^2 - \Sigma_{\ssV\ssV}^{(1)}
\lpar m^2\,,\,m^2\,,\dots\,\rpar + \dots
\eq
\begin{table}[ht]
\caption[]{One-loop $s_{\ssH}(\mh^{\ssO\ssS})$}
\label{of}
\setlength{\arraycolsep}{\tabcolsep}
\renewcommand\arraystretch{1.2}
\vspace{0.3cm}
\begin{tabular}{|l|l|l|}
\hline 
&& \\
$\mh^{\ssO\ssS}\,$ [GeV] & $ 120$ & $300$  \\
&& \\
\hline
$\mu_{\ssH}\,$ [GeV]    & $119.96$ & $299.74$ \\
\hline
$\gamma_{\ssH}\,$ [GeV]  & $7.00\,\times\,10^{-3}$   & $7.90$   \\
\hline
\end{tabular}
\end{table}
To improve the quality of the result we consider the relation
\bq
s_{\ssV} =  m^2 - 
\Sigma_{\ssV\ssV}\lpar s_{\ssV}\,,\,m^2\,,\,\{p\}\,,\,\dots\,\rpar 
\eq
where with $\{p\}$ we denote additional, renormalized, parameters. We then
solve REs and obtain $m^2$ and $\{p\}$, i.e. 
\bq
m^2\,,\,\{p\} = \Reb\,f \lpar s_{\ssV_1}\,,\,s_{\ssV_2}\,,\,\dots\,\rpar,
\eq
where $s_{\ssV_i}$ are {\em experimental} complex poles. Note that we never
expand functions depending on complex POs which means that we actually
compute two-loop diagrams on the second Riemann sheet. Furthermore, in our
scheme the solution of REs is used at the Born level in two-loop diagrams, 
one-loop in one-loop diagrams, two-loop in tree diagrams. 
If $V \not\in \{V_1\,,\,V_2\,,\,\dots\,\}$ we have a genuine prediction,
otherwise we have a consistency relation for loop corrections.
Numerical results for the two-loop $s^{\rm th}_{\ssH}(s^{\rm exp}_{\ssH})$
are shown in \tabn{cf}.

\begin{table}[hb]
\caption[]{Two-loop $s^{\rm th}_{\ssH}(s^{\rm exp}_{\ssH})$. All entries
in GeV.} 
\label{cf}
\setlength{\arraycolsep}{\tabcolsep}
\renewcommand\arraystretch{1.2}
\vspace{0.3cm}
\begin{tabular}{|l|l|l|l|}
\hline 
&&& \\
$\mu_{\ssH}$ & $\gamma_{\ssH}$ & $M_{\ssH}$ & $\Gamma_{\ssH}$ \\
&&& \\
\hline
$300$ & $4$ & $299.96$ & $8.374$ \\
\hline
$300$ & $12$ & $299.87$ & $8.376$ \\
\hline
$500$ & $40$ & $500.17$ & $63.37$ \\
\hline
$500$ & $80$ & $500.42$ & $63.34$ \\
\hline
\end{tabular}
\end{table}
\subsection{Numbers $\&$ renormalization}
Consider one of the REs, e.g.
\bq
\frac{\gf}{\srt} = \frac{g^2}{8\,M^2}\,(1 + \Delta g),
\quad
\Delta g = \delta_{\ssG} + \Delta g^{\ssS},
\label{eqnref:1}
\eq
relating $g$ to the Fermi constant $\gf$; $\Delta g^{\ssS}$ is the $\wb$
self-energy part.
A solution in perturbation theory starts with
\bqa
g^2 &=& 8\,\gf\,\rpw\,\Bigl[ 1 + C^{(1)}_g\,
\frac{\gf}{\pi^2} + \dots \Bigr],
\nl
C^{(1)}_g &=& \frac{1}{2}\,\Bigl[ \Reb\,\Sigma^{(1)}_{\ssW\ssW}(\cpw) -
\Sigma^{(1)}_{\ssW\ssW}(0)\Bigr].
\eqa
According to an old result $\delta^{(1)}_{\ssG}$ is UV/IR finite; we can add 
that $\delta^{(2)}_{\ssG}$ is finite after one-loop renormalization. 
Furthermore, we define a process independent Fermi coupling at the two-loop
level
\bqa
G &=& \gf\,\Bigl\{ 1 - \delta^{(1)}_{\ssG}\,\frac{\gf\,\rpw}{2\,\pi^2}
+ \Bigl[ 2\,(\delta^{(1)}_{\ssG})^2 
\nl
{}&-& \frac{2}{\rpw}\,\delta^{(1)}_{\ssG}\,C^{(1)}_g 
- \delta^{(2)}_{\ssG}\Bigr]\,\lpar \frac{\gf\,\rpw}{2\,\pi^2}
\rpar^2 \Bigr\}.
\eqa
\begin{table}[t]
\caption[]{Percentage two-loop corrections in RE~\eqn{eqnref:1}.}
\setlength{\arraycolsep}{\tabcolsep}
\renewcommand\arraystretch{1.2}
\vspace{0.3cm}
\begin{tabular}{|l|l|l|l|}
\hline 
$\mh^{\ssO\ssS}\,$ [GeV] & $150$ & $300$ & $500$ \\
\hline
&&& \\
$\frac{\gf\,\rpw}{2\,\pi^2}\,
\frac{\delta^{(2)}_{\ssG}}{\delta^{(1)}_{\ssG}}$ &
$18.29\,\%$ & $8.89\,\%$ & $-24.62\%$ \\
&&& \\
\hline
\end{tabular}
\label{rnI}
\end{table}
Therefore, starting from a RE like
\bqa
X &=& x\,( 1 + a_1\,x + a_2\,x^2),
\nl
X &=& \frac{\gf\,\rpw}{2\,\pi^2}, \quad x = \frac{g^2}{16\,\pi^2},
\quad a_1 = \delta^{(1)}_{\ssG} + S^{(1)},
\nl
a_2 &=& S^{(1)}\,\Bigl[
\delta^{(1)}_{\ssG} + S^{(1)}\,\Bigr] 
+ \delta^{(2)}_{\ssG} + S^{(2)},
\eqa
we obtain the following solution:
\bq
x = X + X^2\,( b_1 + b_2\,X),
\;\;
S^{(n)} = \frac{\Sigma^{(n)}_{\ssW\ssW}(0)}{\rpw}.
\label{eqnref:2}
\eq
The LO/NLO/NNLO terms are $X,\,b_1 X^2$ and $b_2 X^3$;
results are shown in \tabns{rnI}{rnII} where one can see that perturbation 
theory becomes questionable beyond $350\,$ GeV.
\begin{table}[ht]
\caption[]{Percentage two-loop corrections in RE~\eqn{eqnref:2}.}
\setlength{\arraycolsep}{\tabcolsep}
\renewcommand\arraystretch{1.2}
\vspace{0.3cm}
\begin{tabular}{|l|l|l|l|}
\hline 
&&& \\
$\mh^{\ssO\ssS}\,$ [GeV] & $150$ & $250$ & $350$ \\ 
&&& \\
\hline
NLO/LO $(\%)$ & $+3.31$ & $-2.30$ & $-7.85$ \\
\hline
$b_1$ & $+12.28$ & $-8.51$ & $-29.07$ \\
\hline
$b_2\,X$ & $+0.25$ & $-1.38$ & $-9.26$ \\
\hline
NNLO/NLO $(\%)$ & $+2.06$ & $+16.16$ & $+31.85$ \\
\hline
\end{tabular}
\label{rnII}
\end{table}
\section{Running of $\alpha$}
The role played by the running of $\alpha$ has been crucial in the development 
of precision tests of the SM.
Underneath this concept there is the popular wisdom that universal corrections
are the important ingredient while non-universal ones should be made as small 
as possible; therefore, universal corrections should be linked to a set of 
POs and data should be presented in the language of POs which, in turn,
is connected with resummation, against gauge invariance. Admittedly, around
$\mz$ it has been easy to perform a discrimination, relevant vs. irrelevant 
terms, paying a little price to gauge invariance. Well above this scale the
situation is drastically different.
Thus, the natural question is about the definition of the running of $\alpha$ 
at an arbitrary scale.
One (fuzzy) idea is to import from QCD the concept of $\MSB$ couplings
and to express theoretical predictions through $\MSB$ 
couplings~\cite{Degrassi:2003rw}. This idea is open for criticism: although 
the $\MSB$ parameter seems unambiguous it violates decoupling.

There is another, well-known, solution: do the calculation in the $R_{\xi}$
gauge, select a $\xi$ - independent part of self-energies, perform resummation
while leaving the rest to ensure independence when combined with vertices and
boxes. The obvious criticism is: it violates uniqueness; however, it is only
a matter of conventions.

Ingredients for $\alpha_{\MSB}$ are: a bosonic part,
a fermionic part with $3$ lepton generations, a perturbative quark 
contribution (top or diagrams where light quarks are coupled internally 
to massive vector bosons) and a non-perturbative one with diagrams where a 
light quark couples to a photon (related to 
$\Delta \alpha^{5}_{\rm had}(\mzs)$). We define
\bqa
\alpha^{-1}_{\MSB}(s) &=& \alpha^{-1} - 
\frac{1}{4\,\pi}\,\Pi^{\MSB}_{\ssQ\ssQ}(0)\bmid_{\mu^2 = s}.
\eqa
Alternatively, we consider $\xi = 1$ and define
\bqa
\frac{\alpha}{\alpha(s)} = 1 + \Delta \alpha(s)
\eqa
Numerical results are shown in \tabn{ralpha}.
\begin{sidewaystable}
\caption[]{The $\MSB$ coupling $\alpha^{-1}_{\MSB}$
and $\Delta \alpha(s)$ at $200\,$GeV.}\vspace{0.2cm}
\begin{tabular}{|l|l|l|l|l|l|}
\hline 
$m_t = 174.3\,\GeV$ & $\mh = 150\,\GeV$ &&&& \\
\hline 
\hline 
{$\sqrt{s}\,\,[\GeV]$} & $\mz$ & $120$ & $160$ & $200$ & $500$ \\
\hline 
one-loop & $128.105$ & $127.974$ & $127.839$ & $127.734$ & $127.305$ \\
two-loop & $128.042$ & $127.967$ & $127.891$ & $127.831$ & $127.586$ \\
$\%$     &           &           &           &           & $0.22$    \\
\hline 
$m_t = 179.3\,\GeV$ & $\mh = 150\,\GeV$ &&&& \\
\hline 
one-loop & $128.113$ & $127.982$ & $127.847$ & $127.742$ & $127.313$ \\
two-loop & $128.048$ & $127.980$ & $127.911$ & $127.857$ & $127.636$ \\
$\%$     &           &           &           &           & $0.25$    \\
\hline 
$m_t = 174.3\,\GeV$ & $\mh = 300\,\GeV$ &&&& \\
\hline 
one-loop & $128.105$ & $127.974$ & $127.839$ & $127.734$ & $127.305$ \\
two-loop & $128.041$ & $127.914$ & $127.784$ & $127.683$ & $127.266$ \\
$\%$     &           &           &           &           & $0.03$    \\
\hline 
\hline
\end{tabular}
\hspace{2.cm}
\begin{tabular}{|l|l|}
\hline 
$\Delta\alpha$ & value at $\sqrt{s} = 200\,$GeV \\
\hline
$\Reb\,$EW  &   $-0.003578(8)$ \\
$\Imb\,$EW  &   $+0.002156(8)$ \\
$\Reb\,$p QCD &  $-0.0005522(4)$ \\
$\Imb\,$p QCD &  $+0.0001178(3)$ \\
fin ren     &   $-0.0000977 - 0.0000998\,i$ \\
\hline
$\Reb\,\alpha(s)$    &   $0.0078782(2)$ \\
$\Reb\,\alpha^{-1}(s)$ &   $126.933(4)$ \\
\hline 
\hline
\end{tabular}
\label{ralpha}
\end{sidewaystable}
\section{Virtual infrared corrections}
We have been able to prove~\cite{Passarino:2006gv} that two-loop vertices 
have an integral representation  
\bq
\dcub{k}(\{x\})\frac{1}{A}\,
\Bigl[ \ln\left( 1 + \frac{A}{B} \right)
\; \mbox{or}\;
\li{n}{\frac{A}{B}}\Bigr],
\eq
where $A, B$ are multivariate polynomials in $\{x\}$, the Feynman parameters. 
Two - loop diagrams are always reducible to combinations of integrals of this 
type where the usual monomials that appear in the integral representation of 
Nielsen - Goncharov generalized polylogarithms are replaced by multivariate 
polynomials of arbitrary degree. 
\vspace{0.5cm}

\noindent
Fig. 3 Example of IR divergent two-loop vertex.
\begin{figure}[ht]
\begin{center}
\vspace{-0.8cm}
\begin{picture}(150,75)(0,0)
 \SetWidth{1.5}
 \SetColor{Cyan}
 \Line(70,-17.5)(40,0)                
 \Line(70,17.5)(40,0)                 
 \Line(70,-17.5)(70,17.5)             
 \SetColor{Brown}
 \Line(0,0)(40,0)                     
 \DashLine(128,-53)(100,-35){2}       
 \DashLine(128,53)(100,35){5}         
 \DashLine(100,-35)(70,-17.5){2}      
 \DashLine(100,35)(70,17.5){5}        
 \SetColor{Red}
 \Photon(100,-35)(100,35){2}{10}
 \SetColor{Black}
 \Text(0,5)[cb]{$-P$}
 \Text(138,-65)[cb]{$p_1$}
 \Text(138,57)[cb]{$p_2$}
 \Text(53,-21)[cb]{$1$}
 \Text(53,14)[cb]{$2$}
 \Text(77,-3)[cb]{$3$}
 \Text(82,-43)[cb]{$m$}
 \Text(82,35)[cb]{$M$}
 \Text(35,50)[cb]{$V^{\bca}_a$}
\end{picture}
\end{center}
\end{figure}

\vspace{1.cm}
\noindent
Our method is fully multi-scale, it allows for a classification of 
infrared divergent configurations, for the evaluation of IR residues and IR 
finite parts and is also suitable for collinear regions. To reach these
objectives we had to extend Berstein - Sato - Tkachov functional 
relations~\cite{Tkachov:1997wh} to higher order transcendental functions.
Results for the diagram in Fig.~3 are shown in \tabn{Cdk}.
\begin{table}[ht]
\setlength{\arraycolsep}{\tabcolsep}
\renewcommand\arraystretch{1.2}
\caption[]{Comparison with the results of 
Davydychev - Kalmykov~\cite{Davydychev:2003mv}. 
Only the infrared finite part is shown, in units of $10^{-8}\,\GeV^{-4}$.}
\vspace{0.3cm}
\begin{tabular}{|l|c|l|l|}
\hline 
 {} & {$\sqrt{s}$} & {$\Reb\,V_{\bcan{0}}$} & 
{$\Imb\,V_{\bcan{0}}$} \\
\hline 
 { Our} & $ 400$  & {$5.1343(1)$} & {$1.94009(8)$} \\
     { DK}  &         &      $5.13445  $  &      $1.94008$  \\
\hline 
 { Our} & $ 300$  & {$5.68801$} & {$-1.61218$} \\
     { DK}  &         &      $5.68801$    &      $-1.61218$  \\
\hline 
 { Our} & $ 200$  & {$9.36340$} & {$-2.84232$} \\
     { DK}  &         &      $9.36340$    &      $-2.84232$  \\
\hline 
 { Our} & $ 100$  & {$29.4726$} & {$-9.74218$} \\
     { DK}  &         &      $29.4726$    &      $-9.74218$  \\
\hline 
\end{tabular}
\label{Cdk}
\end{table}





\noindent
Figure 1. Example of skeleton expansion.
\vspace{-1.cm}

\hspace{-6.2cm}
\begin{minipage}[t]{3.cm}
\vspace{0.1cm}

\includegraphics[height=10.cm,width=9.cm,angle=0,scale=1.3]{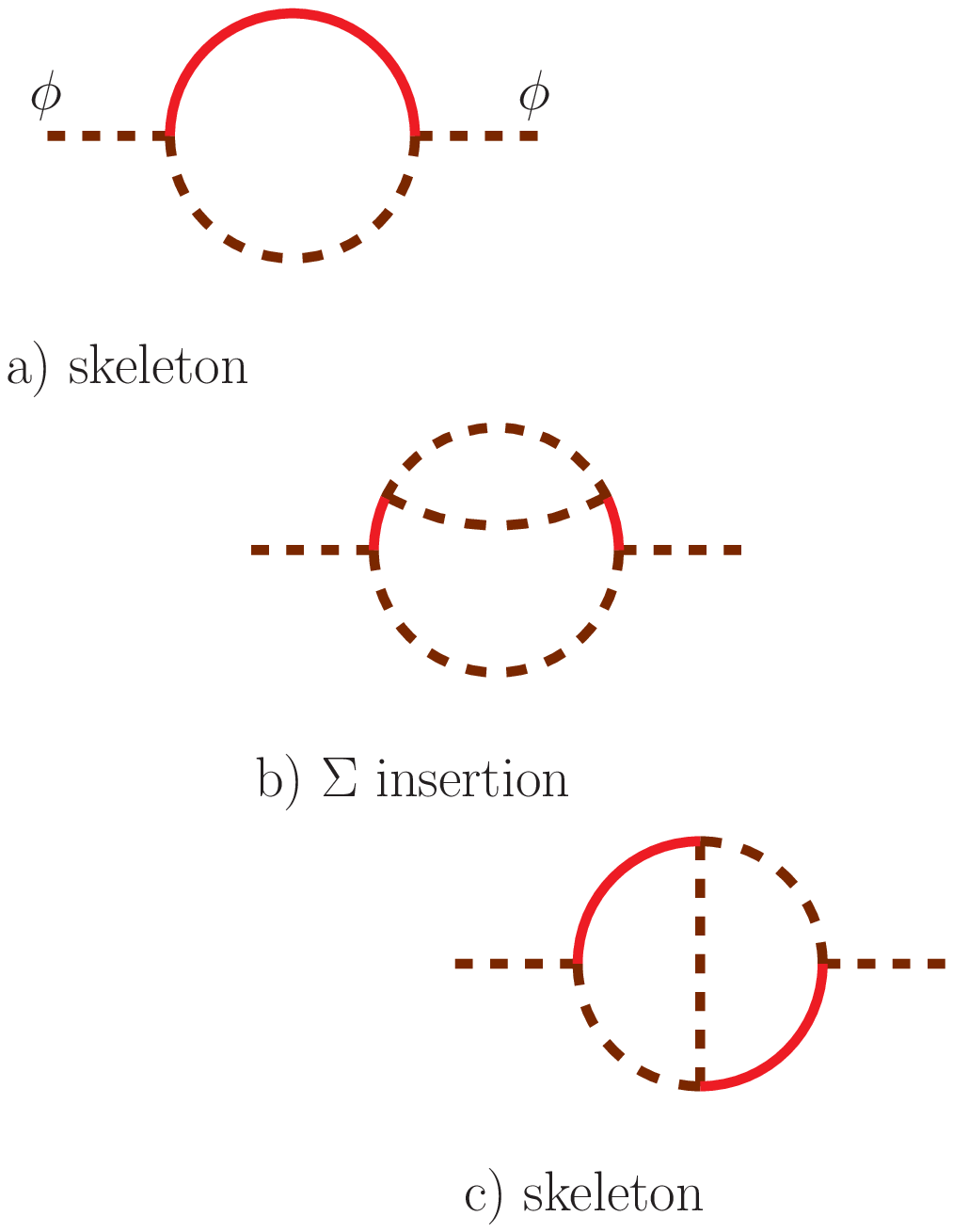}
\end{minipage}

\vspace{-5.8cm}
\noindent
Figure 2. Rearranging perturbation theory in the presence of complex poles.
\vspace{0.7cm}

\hspace{-2.5cm}
\begin{minipage}[b]{3.cm}
\vspace{-1.4cm}

\includegraphics[height=10.cm,width=9.cm,angle=0,scale=1.3]{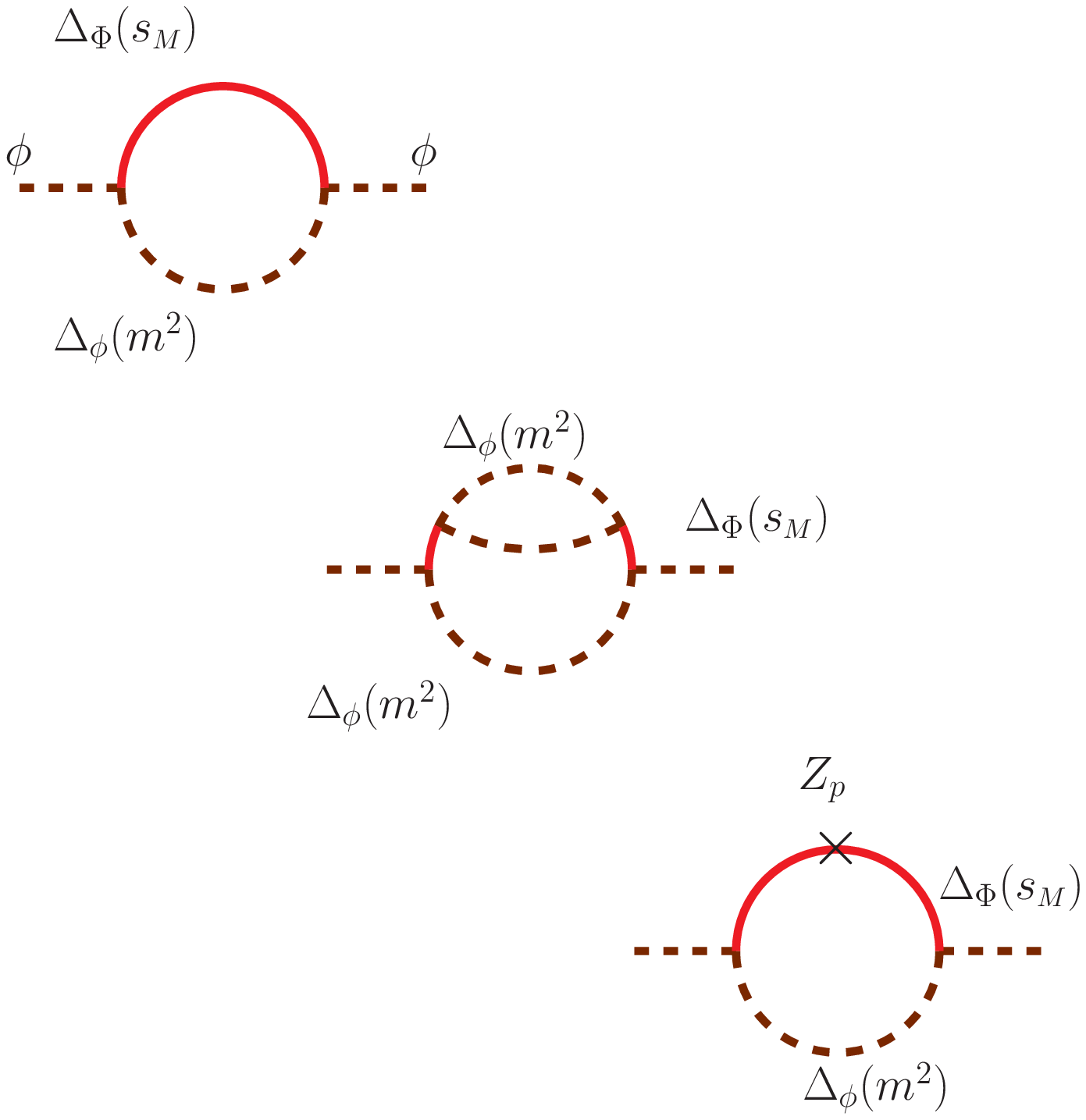}
\end{minipage}

\end{document}